\documentclass[12pt]{article}
\usepackage{etex}
\usepackage{epsfig}
\usepackage{epsf}
\usepackage{latexsym}
\usepackage{amsfonts,amssymb,amsmath} 
\usepackage{mathtools}
\usepackage{color}
\usepackage[a4paper,vmargin=3cm,hmargin=2.1cm]{geometry}
\usepackage{tikz}
\epsfverbosetrue
\newcommand{\de}{\hbox{\rm{d}}}

\newcommand{\pa}{\partial}

\newcommand{\lb}{\left[}
\newcommand{\rb}{\right]}
\newcommand{\lp}{\left(}
\newcommand{\rp}{\right)}

\newcommand{\dpp}{\vcentcolon}
\newcommand{\bb}{\begin{eqnarray}}
\newcommand{\ee}{\end{eqnarray}}
\newcommand{\eee}{\nonumber\end{eqnarray}}
\newcommand{\qq}{\quad}

\begin{document}

\thispagestyle{empty}

\begin{center}
${}$
\vspace{3cm}

{\Large\textbf{Einstein-Cartan, Bianchi I and the Hubble Diagram}} \\

\vspace{2cm}

{\large
Sami R. ZouZou\footnote{ 
Laboratoire de Physique Th\'eorique, D\'epartement de Physique, Facult\'e des Sciences Exactes,\\\indent\qq
Universit\'e Mentouri-Constantine, Algeria
\\\indent\qq
szouzou2000@yahoo.fr
 },
Andr\'e Tilquin\footnote{CPPM, Aix-Marseille University, CNRS/IN2P3, 13288 Marseille, France
\\\indent\qq supported by the OCEVU Labex (ANR-11-LABX-0060) funded by the
"Investissements d'Avenir" 
\\\indent\qq
French government program
\\\indent\qq
 tilquin@cppm.in2p3.fr },
Thomas Sch\"ucker\footnote{
CPT, Aix-Marseille University, Universit\'e de Toulon, CNRS UMR 7332, 13288 Marseille, France
\\\indent\qq supported by the OCEVU Labex (ANR-11-LABX-0060) funded by the
"Investissements d'Avenir" 
\\\indent\qq
French government program
\\\indent\qq
thomas.schucker@gmail.com }
}

\vspace{3cm}

{\large\textbf{Abstract}}

\end{center}
We try to solve the dark matter problem in the fit between theory and the Hubble diagram of supernovae by allowing for torsion via Einstein-Cartan's gravity and for anisotropy via the axial Bianchi I metric.  Otherwise we are conservative and admit only the cosmological constant and dust. The failure of our model is quantified by the relative amount of dust in our best fit: $\Omega_{m0}= 27\,\%\ \pm\ 5 \,\%$ at 1$\sigma $ level.

\vspace{3.cm}

\vskip 1truecm

\section{Introduction}

Today's precision measurements of the Cosmic Microwave Background (CMB) tell us that our universe deviates from maximal space-symmetry at the $10^{-5}$ level. One geometric model for such deviations, based on the Bianchi I metric, was already proposed in the sixties of last century, see Saunders \cite{Sa} and earlier references therein. First signals for eccentricity at 1$\sigma $ level have been found in 2014 by fitting the Bianchi I metric, which solves the Einstein equation with cosmological constant and dust, to the CMB data \cite{cea} and to the Hubble diagram \cite{stv}.

Another signal, equally weak, was found when fitting  the Robertson-Walker metric in the Einstein-Cartan theory \cite{car,hehl76}  (\cite{tworecent} for two recent reviews) to the Hubble diagram: There, torsion generated by some half-integer spin density can be an alternative to dark matter within today's error bars \cite{ts,sz}: $\Omega_{m0}= 9^{+30}_{-\ 7}\,\%$ at 1$\sigma $ level.

It seems natural to try and combine these two approaches. Indeed the axial Bianchi I metric has one privileged direction in the sky in which the spin density may want to point.
     
\section{Dynamics}

The axial Bianchi I metric, $\de \tau^2=\de t^2 - a^2\,(\de x^2+\de y^2)-c^2\, \de z^2,$ has two (positive) scale factors, $a(t)$ and $c(t)$. It is characterised by four Killing vectors, $\pa/\pa x,\,\pa/\pa y,\,\pa/\pa z$ and $x\,\pa/\pa y-y\,\pa/\pa x$, generating three translations and a rotation around the $z$ axis. We want the connection $\Gamma $ to be an independent field. However we want it to respect these four Killing vectors,
\bb 
\xi ^\alpha \,\frac{\pa}{\pa x^\alpha }\,  {\Gamma ^\lambda }_{\mu \nu}
-\,\frac{\pa \xi ^{\lambda  }}{\pa x^{\bar\lambda } }\,  {\Gamma ^{\bar\lambda }}_{\mu \nu}
+\,\frac{\pa \xi ^{\bar\mu }}{\pa x^\mu }\,  {\Gamma ^\lambda }_{\bar\mu \nu}
+\,\frac{\pa \xi ^{\bar\nu }}{\pa x^\nu }\,  {\Gamma ^\lambda }_{\mu \bar\nu}
+\,\frac{\pa ^2\xi ^\lambda }{\pa x^\mu \pa x^\nu}\, 
=0,\label{kill2}\ee
where the $\xi^\alpha,\ \alpha =t,\,x,\,y,\,z $, denote the four components of any of these four Killing vectors. We also want the connection to be metric,
\bb \,\frac{\pa}{\pa x^\lambda }\, g_{\mu \nu}
-{\Gamma ^{\bar\mu  }}_{\mu \lambda } g_{\bar\mu \nu}
-{\Gamma ^{\bar\nu }}_{\nu \lambda } g_{\mu\bar \nu}=0.\label{metric}\ee
The most general metric connection invariant under the three translations and the rotation involves eight arbitrary functions of time, four of which are parity even and the other four are odd. Since the Bianchi I metric is invariant under space inversion we also want the connection to be parity even. Then it can have at most the following non-vanishing components:
\begin{align}
{\Gamma ^x}_{xt}&=a'/a,&
{\Gamma ^x}_{tx}&=b/a,&
{\Gamma ^t}_{xx}&=ab,&
{\Gamma ^x}_{ty}&=+\,g,\ \,{}&
{\Gamma ^x}_{yt}&=+\,h,\ \,{}
\\[2mm]
{\Gamma ^y}_{yt}&=a'/a,&
{\Gamma ^y}_{ty}&=b/a,&
{\Gamma ^t}_{yy}&=ab,&
{\Gamma ^y}_{tx}&=-\,g,\ \,{}&
{\Gamma ^y}_{xt}&=-\,h,\ \,{}
\\[2mm]
{\Gamma ^z}_{zt}&=c'/c,&
{\Gamma ^z}_{tz}&=d/c,&
{\Gamma ^t}_{zz}&=cd,&
{\Gamma ^t}_{xy}&=+a^2g,&
 {\Gamma ^t}_{yx}&=-a^2g,
\end{align}
with four arbitrary functions of time: $b,\,d,\,g$ and $h.$ To simplify our analysis we eliminate three of these four functions by choosing: $b=a'$, $d=c'$ and $h=-g$. To justify this choice, we note that it entails a completely antisymmetric torsion. We like completely antisymmetric torsion for two reasons:
 
{$\,\,$\it (i)} It has connection geodesics that automatically coincide with metric geodesics.

{\it (ii)} It can be generated by Dirac spinors.\\
So far we have expressed our connection in a holonomic frame $\de x^\mu $ where we denoted it by $\Gamma $. It will be convenient to express it in an orthonormal frame $e^a={e^a}_\mu\, \de x^\mu $ where we denote it by $\omega $. In components we have:
\bb {\omega ^a}_{b \mu }={e^a}_\alpha\,  {\Gamma^\alpha  }_{\beta \mu }\,{e^{-1\,\beta }}_b +{e^a}_\alpha\,\frac{\pa}{\pa x^\mu }\,   {e^{-1\,\alpha  }}_b.\label{gaugetrans}\ee
In an orthonormal frame the metricity condition (\ref{metric}) is purely algebraic and states that ${\omega ^a}_{b}\dpp={\omega ^a}_{b\mu }\,\de x^\mu $ is a 1-form with values in the Lorentz algebra: 
\bb
{\omega }_{ab\mu }=-{\omega }_{ba\mu },\ee
where orthonormal (latin) indices are raised and lowered with\\ 
$\eta^{ab } =\eta_{ab } =\,\,$diag$\,(1,\,-1,\,-1,\,-1)$.

Of course we choose ${e^a}_\mu =\,\,$diag$\,(1,\,a,\,a,\,c)$. Then the non-vanishing components of ${\omega ^a}_{b\mu }$ are:
\begin{align}
{\omega ^t}_{xx }={\omega ^x}_{tx }={\omega ^t}_{yy }
={\omega ^y}_{ty}&=a',& {\omega ^t}_{xy }={\omega ^x}_{ty }&=ag,&
{\omega ^t}_{yx }={\omega ^y}_{tx }&=-ag,&
\\
{\omega ^t}_{zz }={\omega ^z}_{tz}&=c',
&
{\omega ^x}_{yt }&=-g,&{\omega ^y}_{xt }&=g.
\end{align}
The torsion 2-form $T_a\dpp=\eta_{a\tilde a}\,(\de e^{\tilde a}+{\omega ^{\tilde a}}_{d }\,e^d)
=\dpp {\textstyle\frac{1}{2}} T_{abc}\,e^b\,e^c$ has completely antisymmetric components:
\bb T_{txy}=T_{xyt}=T_{ytx}=-2g.\ee
The curvature 2-form ${R^a}_b\dpp=\de\,{\omega ^a}_b+
{\omega ^a}_f{\omega ^f}_b=\dpp{\textstyle\frac{1}{2}}  {R^a}_{bcd}\,e^c\,e^d$ takes values in the Lorentz algebra and has the following components:
\begin{align}
{R^t}_{xtx}&={R^t}_{yty}=\,\frac{a''}{a}\, +g^2,
&
{R^t}_{ztz}&=\,\frac{c''}{c}\, ,
&
{R^t}_{xty}&=-{R^t}_{ytx}=g',
\\[2mm]
{R^x}_{yxy}&=\lp\frac{a''}{a}\rp^2+g^2,
&
{R^{x}}_{zxz}&={R^{y}}_{zyz}=\,\frac{a'}{a}\,\frac{c'}{c}\, ,
&
{R^z}_{xzy}&=-{R^z}_{yzx}=\,\frac{c'}{c}\,g.
\end{align}
For the Ricci tensor ${\rm Ric\,}_{bd}\dpp ={R^a}_{bad}$ we find:
\bb 
{\rm Ric\,}_{tt}=-2\,\frac{a''}{a}\, -\,\frac{c''}{c}\, -\,2\,g^2,&&
{\rm Ric\,}_{xy}=-{\rm Ric\,}_{yx}=g'\,+\,\frac{c'}{c}\,g,
\\[2mm]
{\rm Ric\,}_{xx}={\rm Ric\,}_{yy}=\,\frac{a''}{a}\,+\lp \frac{a'}{a}\rp^2+\,\frac{a'}{a}\, \frac{c'}{c}\,+\,2\,g^2,
&&
{\rm Ric\,}_{zz}=\,\frac{c''}{c}\,+2\,\frac{a'}{a}\, \frac{c'}{c}\,.
\ee
The curvature scalar is:
\bb 
{\rm R}=
-4\,\frac{a''}{a}\,-\,2\,\frac{c''}{c}\,-2\lp \frac{a'}{a}\rp^2-4\,\frac{a'}{a}\, \frac{c'}{c}\,-\,6\,g^2,
\ee
and the Einstein tensor $G_{ab}\dpp={\rm Ric\,}_{ab}-{\textstyle\frac{1}{2}} \,{\rm R}\,\eta_{ab}$ has components:
\begin{align} 
G_{tt}=\lp \frac{a'}{a}\rp^2+\,2\,\frac{a'}{a}\, \frac{c'}{c}\, +\,g^2,&&
G_{xy}=-G_{yx}=g'\,+\,\frac{c'}{c}\,g,&
\\[2mm]
G_{xx}=G_{yy}=-\,\frac{a''}{a}\, -\,\frac{c''}{c}\,-\,\frac{a'}{a}\, \frac{c'}{c}\,-\,g^2,
&&
G_{zz}=-2\,\frac{a''}{a}\,-\lp \frac{a'}{a}\rp^2-\,3\,g^2.&
\end{align}
Finally we can write down Einstein's equation, $G_{ab}-\Lambda \,\eta_{ab}=8\pi\,G\,\tau_{ba} $ for dust with mass density $\rho $ and $\tau_{ab}={\rm diag}\,(\rho ,\,0,\,0,\,0)$:
\begin{align}
(t&t)& \lp\,\frac{a'}{a}\,\rp ^2+2\,\frac{a'}{a}\,\frac{c'}{c}\,
+g^2-\Lambda& =8\pi\, G\,\rho ,&\\[2mm]
(x&x)&\frac{a''}{a}\,+\,\frac{c''}{c}\,+\,\frac{a'}{a}\,\frac{c'}{c}\,+g^2-\Lambda &=0,\\[2mm]
(z&z)&2 \,\frac{a''}{a}\,+\lp\,\frac{a'}{a}\,\rp^2+3\,g^2-\Lambda &=0,\\[2mm]
(x&y)&\, g'+\,\frac{c'}{c}\,g&=0.
\end{align}
 The last equation integrates to $g=\pm\sqrt{\kappa }\,c_0/c$. Note that the difference of the $xx$ and $zz$ components implies that $g$ must vanish in the isotropic case, $a=c$. However the limit of $\kappa $  as the Hubble stretch $h_{x0}$ (defined below) tends to zero is not continuous.

By Cartan's equation, the parity even and completely antisymmetric torsion is sourced by a spin density $s$ parallel to the $z$ direction and Cartan's equation reduces to $g=4\pi \,G\,s$. 

The first derivative of the $tt$ component yields, upon use of the other components, the covariant conservation of energy:
\bb 8\pi G\lb\rho' +\lp2\,\frac{a'}{a}\, +\,\frac{c'}{c}\, \rp\rho \rb=-4\kappa \,\frac{c_0^2\,c'}{c^3}\, .\ee
Let us use the following notations: the directional and the mean Hubble parameters:
\bb H_x\dpp=\,\frac{a'}{a}\,, \qq H_z\dpp=\,\frac{c'}{c}\,, \qq H\dpp={\textstyle\frac{1}{3}}\,(2H_x+H_z),\ee
 and the Hubble stretches $h_x$ and $h_z$ defined by:
\bb 
H_x=\dpp H\,(1+h_x),\qq H_z=\dpp H\,(1+h_z)=H\,(1-2h_x).\ee
The last equation results from $2h_x+h_z=0$. We also use the dimensionless functions of time:
\bb 
\Omega _\Lambda \dpp=\,\frac{\Lambda }{3H^2}\, ,\qq
\Omega _m \dpp=\,\frac{8\pi G\,\rho  }{3H^2}\, ,\qq
\Omega _\kappa  \dpp=\,\frac{\kappa \,(c_0/c)^2  }{3H^2}\, .\ee
With these we have the following system of 5 ordinary, first order differential equations:
\begin{align}
&&&\,\frac{a'}{a}\, = H\,(1+h_x),\label{equ1}\\[2mm]
&&&\,\frac{c'}{c}\, = H\,(1-2h_x),\label{equ2}\\[2mm]
(xx)&&& 2H'-H'h_x-H\,h'_x+3H^2[1-h_x+h_x^2]=3H^2\lp
\Omega _{\Lambda 0}\,\frac{H_0^2}{H^2}\, -\Omega _{\kappa 0}
\,\frac{c_0^2H_0^2}{c^2H^2}\, \rp,\label{equ3}\\[2mm]
(zz)&&& 2H'+2H'h_x+2H\,h'_x+3H^2[1+2h_x+h_x^2]=3H^2\lp
\Omega _{\Lambda 0}\,\frac{H_0^2}{H^2}\, -3\,\Omega _{\kappa 0}
\,\frac{c_0^2H_0^2}{c^2H^2}\, \rp\!,\label{equ4}\\[2mm]
{\rm (cons.)}&&& \Omega _m'+\lb2\,\frac{H'}{H}\, +3\,H\rb\Omega _m=-4H\,(1-2h_x)\,\Omega _{\kappa 0}
\,\frac{c_0^2H_0^2}{c^2H^2}\, ,\label{equ5}
\end{align}for 5 unknown functions of time: $a,\,c,\,H,\,h_x, \Omega _m$; with 5 initial conditions: $a_0=1,\,c_0=1, H_0,\,h_{x0},\,\Omega _{m0}$ and one external parameter: $ \Omega _{\kappa 0}$.
Note that the initial conditions $a_0$ and $c_0$ can be chosen arbitrarily (positive) and by the $tt$ component of Einstein's equation today we have 
\bb
\Omega _{\Lambda 0}=1-\Omega _{m0}+\Omega _{\kappa 0}-h_{x0}^2.
\ee
We were unable to solve the system analytically and asked Runge \& Kutta for their kind help. We checked their numerical solution in the torsionless case, $\kappa =0$,  against the known analytical solution \cite{stv}.

\section{Kinematics}

The Hubble diagram is the parametric plot of the apparent luminosity $\ell$ as a function of redshift $z$ and as a function of the direction  $( \theta ,\,\varphi )$ of the incoming photons emitted at time $t_{-1}( \theta ,\,\varphi )$ by a Super Nova Ia. As the emission time is not observable, this is the parameter to be eliminated. The absolute luminosity $L$ of all Super Novae Ia is supposed to be the same. 

As the torsion is completely antisymmetric, it modifies the dynamics of the metric but not the geodesics. Therefore we can take the formulas for direction, redshift and apparent luminosity from the literature \cite{Sa,stv}:

Let us denote the unit vector pointing towards the Super Nova by
\begin{align}
\qq\qq\qq\qq\qq A&\dpp=a_0\,\cos\varphi  \,\sin\theta  ,&\\
B&\dpp=a_0\,\sin\varphi  \,\sin\theta  ,&\\
C&\dpp=a_0\,\qq\qq \,\cos\theta  ,&
\end{align}
and the arrival time of the photons of all Super Novae (today) by $t_0=0$.
Put $a_0\dpp=a(t_0)=a(0)=1,\ a_{-1}\dpp=a(t_{-1})$, ..., and define the function
\bb W(t)\dpp =\lp \,\frac{A^2+B^2}{a(t)^2}\,  +
\,\frac{C^2}{c(t)^2}\, \rp ^{-1/2}.\label{wt}\ee
Then solving the geodesic equations for the photons in an axial Bianchi I metric, we obtain the redshift:
\bb z\dpp=\,\frac{W_0}{W_{-1}}\, -1\,=\,\sqrt{\,\frac{ A^2+B^2}{a^2_{-{1}}}\,+\,\frac{ C^2}{c^2_{-{1}}}\ }\,-1\,, \label{redshift}
 \ee
 and the apparent luminosity:
 \bb \ell=\,\frac{L}{4\pi }\, \frac{  a_0\,W_{-1}^5}{a^2_{-1}c_{-1}\,W_0^4\,\Delta }\, ,\label{lumi}\ee
with:
\bb \Delta \dpp = \lb(A^2+B^2)I_xI_z  +C^2I_x^2\rb/a_0^2,\qq I_x\dpp=-a_0^3\int_{t_{-1}}^{t_0}\frac{W^3}{a^2c^2}\,  \de t,\label{ixiz}
\qq
I_z\dpp=-a_0^3\int_{t_{-1}}^{t_0} \frac{W^3}{a^4}\,\de t.
\ee 
To compute the emission time $t_{-1}(z,\,\theta ,\,\varphi )$ for a given direction and redshift we have to invert the numerical function $W(t_{-1})$ in any fixed direction. Note that there are cases where one of the scale factors goes through a minimum. Then $W(t_{-1})$ is not invertible for certain directions, the apparent luminosity becomes a double-valued function of redshift \cite{non} and the Hubble diagram is a truly parametric plot.

\section{Data analysis}

 This analysis is very similar to \cite{stv}. Only a quick reminder is given
 here.

 To test the Bianchi 1 - Cartan hypothesis we use the type 1a supernovae Hubble
 diagram with the Union 2 data sample \cite{union2} containing 557 supernovae
 up to a redshift of 1.4 and the Joint Light curve Analysis (JLA) \cite{jla}
 data sample containing 740 supernovae up to a redshift of 1.3 and 258 common
 supernovae with Union 2. The celestial coordinates of the Supernovae  are obtained from the SIMBAD database \cite{simbad}. 

For Union 2 sample, the published magnitudes at maximum luminosity are
marginalized over the time stretching of the light curve and the color at maximum
brightness. Statistical and systematical errors on the associated magnitudes are
provided by the full covariance matrices. The JLA published data provide 
the observed uncorrected peak magnitude  ($m_{\rm peak}$), the time stretching ($X1$)
and the color ($C$) with the full statistical and systematical covariance matrices.
The total $\chi^2$ reads:
\bb \chi^2 = \Delta M^T V^{-1} \Delta M, \ee
Where $\Delta M$ is the vector of differences between expected 
and reconstructed magnitudes at maximum and $V$ is the full covariance matrix. 


The expected magnitude is written as $m_e(z) = m_s - 2.5 \log \ell(z)$ 
where $m_s$ is a normalization parameter. For the combined sample (1007
supernovae), we use two different normalization parameters (one for each
sample) because of the different  light curve calibrations.
 
 The apparent luminosity  $\ell$  is computed in two
 steps. For each set of cosmological parameters, we solve the differential equations
 (\ref{equ1}-\ref{equ5})  by the help of a
Rugge Kutta algorithm during the minimization
 procedure.  To keep numerical
instabilities low we choose a time step of 1000 years. The two scale factors, $a(t)$ and $c(t)$, are stored in the memory
for the next step. Then, for a given privileged direction and for each supernova
at a given redshift and angles $\theta$ and $\varphi $ with respect to privileged
direction, we scan the emission time
of the photon with equation (\ref{redshift}) up to the supernova redshift. 
At the same time we compute both integrals $I_x$, $I_z$ (\ref{ixiz}) by a
direct summation using formula (\ref{wt}) and the corresponding scale factors 
from the previous step. Finally we compute the apparent luminosity by the use of (\ref{lumi}).

We use the MINUIT package \cite{minuit} to minimize the $\chi^2$.  We choose
 the SIMPLEX algorithm known as Nelder-Mead method \cite{simplex} for
minimization, because it is well suited for nonlinear optimization problems without knowledge of derivatives.
The numerical iterative minimization is thus less sensitive to numerical instabilities.

We search for privileged directions by scanning the celestial sphere in step
of $4\times 4$ square degrees in right ascension and declination to keep the
computing time  within reasonable amounts, more than $200\,000$ hours on a single CPU. For each direction
we minimize the $\chi^2$ with respect to the cosmological parameters
$\Omega_{m0}$, $\Omega _{\kappa 0}$, $h_{x0}$ and the nuisance parameter
$m_s$.

\begin{figure}[h]
\begin{center}
\epsfig{figure=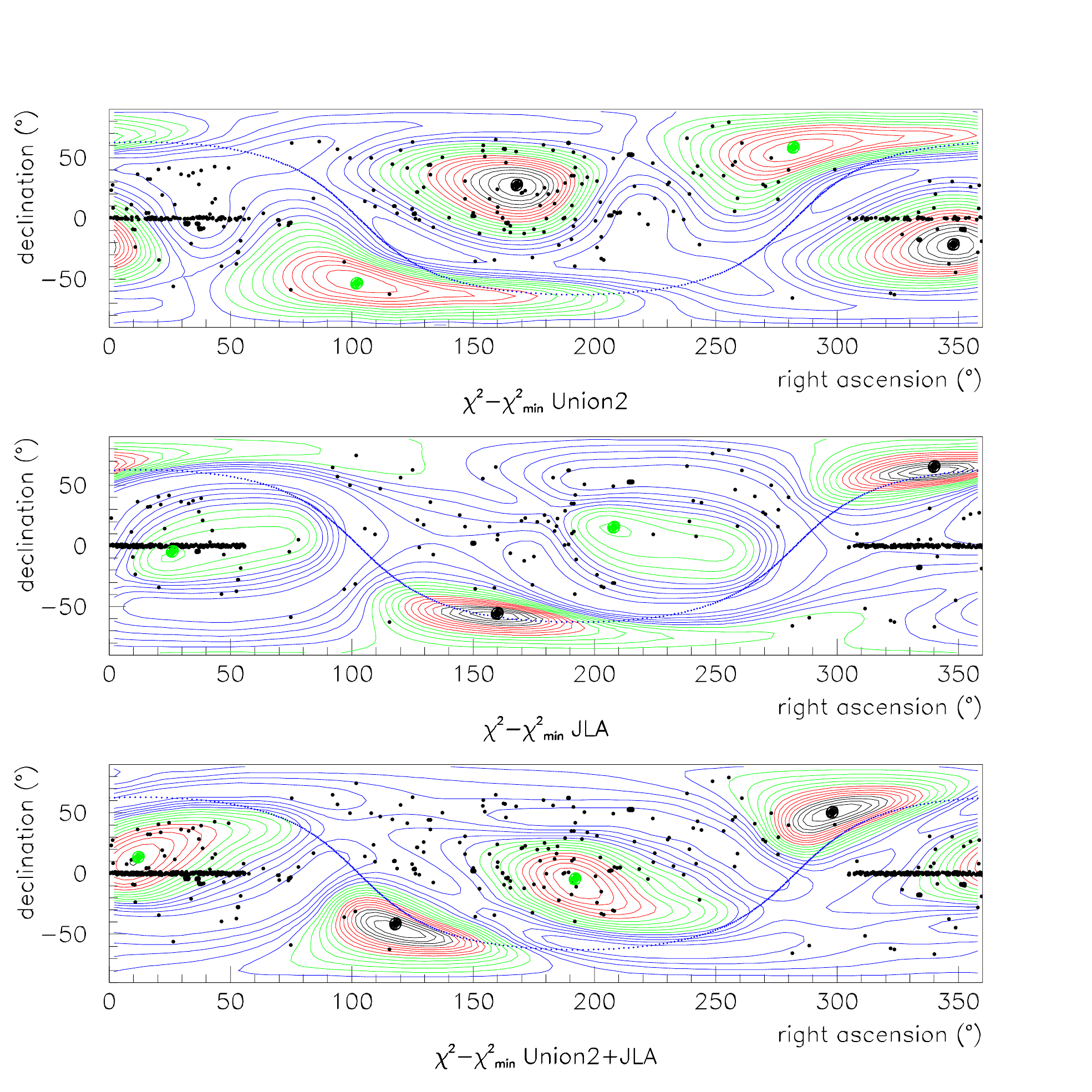,width=15cm}
\caption[]{Confidence level contours of the privileged directions in arbitrary color
  codes for Bianchi I Einstein-Cartan spacetimes. Black points
  represent supernova positions.
  Note the accumulation  of supernovae in the equatorial plane. The blue line is the
  galactic plane. Black specks represent the main privileged direction
  while the green specks show the secondary direction.}
\label{fig1}
\end{center}
\end{figure}

 Figure \ref{fig1} shows the confidence level contours in arbitrary
color codes around the eigen-directions on the celestial sphere for  Union 2,
JLA and the combined sample. Smooth contours have been obtained by the use of a
Multi Layer Perceptrons (MLP) neural network \cite{frank}\cite{neural} with 2 hidden layers of 40 neurons each trained on
the results of the fit. Black points in figure \ref{fig1} shows the supernovae
positions, black (respectively green) specks the main (respectively secondary)
direction. Notice the back-to-back symmetry due to the space reflexions symmetry
of the Bianchi I metric.
 Compared to figure 1 in \cite{stv}, the contours are very simular exept
that the main direction for Union 2 becomes the secondary one. This is not
unexpected because the models are different and statistical significance is only
1$\sigma$. Furthermore, the Hubble stretch $h_{x0}$ has the opposite
sign (Table \ref{table1}).

Table \ref{table1} shows our results of the fit and the 1$\sigma$ error. As
the SIMPLEX algorithm does not provide any errors, we compute them for each
parameter $p_i$ by solving numerically the equation $\chi^2(p_i) = \chi^2_{\rm min} + 1$
where $\chi^2_{\rm min}$ is the minimum of $\chi^2$ on all others parameters. To do
that we scan each parameter around the minimum and interpolate $\chi^2(p_i)$
by a polynomial function to smooth the numerical instabilities and estimate
the 1$\sigma$ error. 

Comparing our results with those of Table 1 in \cite{stv}, we find them 
statistically compatible. The present minimum $\chi^2$ is better by about
one unit which is expected with one more degree of freedom, $h_{x0}$.
For all three samples, the added  parameters describing torsion, $\Omega
_{\kappa 0}$, and anisotropy, $h_{x0}$, are statistically compatible
with zero. Therefore supernovae alone are not accurate enough
to detect these two features. 

The last column of Table \ref{table1} shows expected errors for one year of LSST. The speed up
the final analysis we
randomly simulate only one tenth of the total expected number of supernovae
for one year of LSST (50000) in 20000 square degrees with redshift 
between 0.4 and 0.8. As fiducial cosmology we take the JLA experimental results.
For the analysis we use a magnitude error of 0.12 and a
redshift error of $0.01(1+z)$ propagated to the magnitude error. The final errors
on the magnitude are then scaled down by $\sqrt{10}$ in order to emulate the 1
year LSST statistic. No systematic errors have been used and we did not try
to compute contours, which would have required about 1 million hours of CPU. 
As shown in Table \ref{table1} the improvement on the errors scales down
approximately as the square root of the number of supernovae.

In the future, this analysis would have to be improved to deal with numerical
instabilities and computing time. We could either
linearize the solution of the differential equations or if not possible use a
neural network to parameterize the supernovae luminosity with respect to
$\theta$, $\Omega _{m0}$, $\Omega _{\kappa 0}$, $h_{x0}$ and redshift.

\begin{table}[htbp]
\begin{center}
\begin{tabular}{||c||c|c|c|c||} \hline
                    &   Union II      &  JLA     &     Combined   &       LSST (1 year) \\  \hline
ascension($^\circ$)      &$168  \pm 33$ &$ 161 \pm 42$ &$ 120 \pm 50$    &  $ \pm 4$  \\  \hline
declination($^\circ$)    &$ 26 \pm 36$  &$ -49 \pm 20$ &$ -36 \pm 36$    &  $     \pm 2$     \\  \hline
$\Omega _{m0} $              &$ 0.24 \pm 0.05$ & $0.27 \pm 0.05 $ & $0.27 \pm  0.05$  & $ \pm 0.002 $ \\  \hline
$\Omega _{\kappa 0}$ &$ 0.02^{+0.05}_{-0.02}$ & $ 0.^{+0.04}_{-0.0} $ &  $0.^{+0.02}_{-0.0}$ & $ \pm 0.002$  \\ \hline
$h_{x0}$            &$0.009 \pm 0.12$ & $ 0.01^{+0.006}_{-0.01}$ &  $0.005^{+0.003}_{-0.005}$ & $\pm 4 \cdot10^{-4}$ \\ \hline
$\chi ^2_{\rm min}$  & $529.4$ & $699.5$ & $1001.7$ &  \\ \hline 

\end{tabular}
\caption[]{1$\sigma$ errors for the Bianchi I Cartan cosmological parameters  
  for Union 2, JLA and combined samples and for one year of LSST.}
\label{table1}
\end{center}
\end{table}

\section{Conclusion}

The cosmological principle allows for
a spin density pointing in the ``time  direction''. In Einstein-Cartan's gravity, this spin density generates a torsion component in the same direction. At 1$\sigma $ confidence level, this torsion component can solve the dark matter problem in the Hubble diagram \cite{ts,sz}. 
Our motivation for this work was the hope that relaxing the cosmological principle and allowing the spin density to point into a privileged direction in the sky could improve this confidence level. 

Today's observational data have clearly decided -- after $200\,000$ CPU hours -- not to support our hope.

${}$\vspace{6mm}

\noindent
{\bf Acknowledgements:} This work has been carried out thanks to the support of the OCEVU Labex
(ANR-11-LABX-0060) and the A*MIDEX project (ANR-11-IDEX-0001-02) funded
by the "Investissements d'Avenir" French government program managed by
the ANR.

\hfil\eject


\begin{thebibliography}{10}
 
\bibitem{Sa} P. T. Saunders, ``Observations in some simple cosmological models with shear," 
Mon. Not. R. Astr. Soc. {\bf 142} (1969) 213.  
 \bibitem{cea}
  P.~Cea,
  ``The Ellipsoidal Universe in the Planck Satellite Era,''
  Mon.\ Not.\ Roy.\ Astron.\ Soc.\  {\bf 441} (2014) 1646
  [arXiv:1401.5627 [astro-ph.CO]].
\bibitem{stv}
  T.~Sch\"ucker, A.~Tilquin and G.~Valent,
  ``Bianchi I meets the Hubble diagram,''
  Mon.\ Not.\ Roy.\ Astron.\ Soc.\  {\bf 444} (2014) 2820
  [arXiv:1405.6523 [astro-ph.CO]].
  \bibitem{car} \'E.~Cartan, {\it Sur les vari\'et\'es \`a connexion affine et la th\'eorie de la r\'elativit\'e g\'en\'eralis\'ee (premi\`ere partie),} Ann.~\'Ec.~Norm.~Sup.~{\bf 40} (1923) 325.\\
{\it (premi\`ere partie, suite),} Ann.~\'Ec.~Norm.~Sup.~{\bf 41} (1924) 1.\\
 {\it (deuxi\`eme partie),} Ann.~\'Ec.~Norm.~Sup.~{\bf 42} (1925) 17.
 \bibitem{hehl76}
 F. W. Hehl, P. von der Heyde, G. D. Kerlick and J. M. Nester, {\it General relativity with spin and torsion: Foundations and prospects,} Rev. Mod. Phys. {\bf 48} (1978) 393.
 \bibitem{tworecent}
 S.~Capozziello, G.~Lambiase, C.~Stornaiolo,
{\it Geometric classification of the torsion tensor in space-time,}
  Annalen Phys.\  {\bf 10 } (2001)  713.
  [gr-qc/0101038].\\
 I.~L.~Shapiro,
 {\it Physical aspects of the space-time torsion,}
  Phys.\ Rept.\  {\bf 357 } (2002)  113.
  [hep-th/0103093].
  \bibitem{ts}
  A.~Tilquin and T.~Sch\"ucker,
  ``Torsion, an alternative to dark matter?,''
  Gen.\ Rel.\ Grav.\  {\bf 43} (2011) 2965
  [arXiv:1104.0160 [astro-ph.CO]].
  \bibitem{sz}
  T.~Sch\"ucker and S.~R.~ZouZou,
  ``On a weak Gauss law in general relativity and torsion,''
  Class.\ Quant.\ Grav.\  {\bf 29} (2012) 245009
  [arXiv:1203.5642 [gr-qc]].
  \bibitem{non}
  T.~Sch\"ucker and A.~Tilquin,
  ``From Hubble diagrams to scale factors,''
  Astron.\ Astrophys.\  {\bf 447} (2006) 413
  [astro-ph/0506457].
  \bibitem{union2}
    R.~Amanullah, C.~Lidman, D.~Rubin, G.~Aldering, P.~Astier, K.~Barbary, M.~S.~Burns and A.~Conley {\it et al.},
 ``Spectra and Light Curves of Six Type Ia Supernovae at $0.511 < z < 1.12$ and the Union2 Compilation,''
  Astrophys.\ J.\  {\bf 716} (2010) 712
  [arXiv:1004.1711 [astro-ph.CO]].
  \bibitem{jla}  
  M.~Betoule {\it et al.} [SDSS Collaboration],
  ``Improved cosmological constraints from a joint analysis of the SDSS-II and SNLS supernova samples,''
  Astron.\ Astrophys.\  {\bf 568} (2014) A22
  [arXiv:1401.4064 [astro-ph.CO]].
  \bibitem{simbad} SIMBAD astronomical database:
    http://simbad.u-strasbg.fr/simbad/  
  \bibitem{minuit} ``The ROOT analysis package,''
    http://root.cern.ch/drupal/
  \bibitem{simplex}
  John Nelder and Roger Mead, `` A simplex method for function minimization '', Computer Journal, vol. 7, no 4 (1965) 308
  \bibitem{frank}
  Frank Rosenblatt, ``A Probabilistic Model for Information Storage and
    Organization in the Brain, ''
    Cornell Aeronautical Laboratory, Psychological Review, vol. 65, no. 6 (1958)  386. 
  \bibitem{neural} ``TMultiLayerPerceptron: Designing and using Multi-Layer
    Perceptrons with ROOT, ''
    http://cp3.irmp.ucl.ac.be/~delaere/MLP/

\end{thebibliography}
\end{document}